\pdfoutput=1
\documentclass{PoS}

\usepackage{amsmath}

\def\la{\langle}
\def\ra{\rangle}

\def\beq{\begin{equation}}
\def\eeq{\end{equation}}
\def\bea{\begin{eqnarray}}
\def\eea{\end{eqnarray}}
\def\barr{\begin{array}}
\def\earr{\end{array}}

\title{Phase structure and real-time dynamics\\of the massive Thirring
model in 1+1 dimensions\\using the tensor-network method\thanks{The authors warmly thank Pochung Chen for discussions.  Numerical simulations
have been carried out on the LOEWE-CSC high-performance computer of Johann
Wolfgang Goethe-University Frankfurt am Main, and on the
high-performance computing
facilities at National Chiao-Tung University.}}

\ShortTitle{Phase structure and real-time dynamics of Thirring model
  from MPS}

\author{Mari~Carmen~Ba\~{n}uls$^{a,b}$\thanks{Paritally supported by the Deutsche Forschungsgemeinschaft (DFG, German Research Foundation) under Germany's Excellence Strategy -- EXC-2111 -- 390814868,
and by the EU-QUANTERA project QTFLAG (BMBF grant No. 13N14780).}\, , \,
  Krzysztof~Cichy$^{c}$\thanks{Supported by National Science Centre (Poland) grant SONATA BIS
2016/22/E/ST2/00013.}\, , \, 
  Hao-Ti~Hung$^{d}$\thanks{Supported in part by
Ministry of Science and Technology (MoST) of Taiwan under Grants No.
105-2112-M-002-023-MY3.}\, , \,
  Ying-Jer~Kao$^{d}$\thanks{Supported in part by
Ministry of Science and Technology (MoST) of Taiwan under Grants No.
105-2112-M-002-023-MY3, 107-2112-M-002 -016 -MY3, and 108-2918-I-002
-032.}\, , \,
  \speaker{C.-J.~David~Lin}$^{\mbox{ }e,f}$\thanks{Partially supported by Taiwanese MoST
Grant No. 105-2628-M-009-003-MY4.}\, , \,
  Yu-Ping~Lin$^{g}$\thanks{Sponsored by the Army
Research Office under Grant No. W911NF-17-1-0482.}\,\,\, , \, David~T.-L.~Tan$^{e}$\thanks{Supported by Taiwanese MoST
Grant No. 105-2628-M-009-003-MY4.}\\
             \llap{$^a$}Max Planck Institut f\"{u}r Quantenoptik, Garching 86748,
        Germany\\
             \llap{$^b$}Munich Centre for Quantum Science and Technology, Munich 80799, Germany\\
              \llap{$^c$}Faculty of Physics, Adam Mickiewicz
              University, ul. Uniwersytetu Pozna\'nskiego 2,\\61-614 Pozna\'{n}, Poland\\
             \llap{$^d$}Department of Physics, National Taiwan University, Taipei 10617
        Taiwan\\
             \llap{$^e$}Institute of Physics, National Chiao-Tung University,
        Hsinchu 30010, Taiwan\\
             \llap{$^f$}Centre for High Energy Physics, Chung-Yuan Christian
        University, Chung-Li, 32032, Taiwan\\
             \llap{$^g$}Department of Physics, University of Colorado,
        Boulder, CO 80309, USA\\
             Email:\email{banulsm@mpq.mpg.de},
             \email{kcichy@amu.edu.pl}, \email{hunghaoti852@gmail.com}, \email{yjkao@phys.ntu.edu.tw}, \email{dlin@mail.nctu.edu.tw}, \email{Yuping.Lin@colorado.edu}, \email{tanlin2013.py04g@nctu.edu.tw}}

\abstract{We present concluding results from our study for
  zero-temperature phase structure
of the massive Thirring model in 1+1 dimensions with staggered
regularisation. Employing the method of matrix product states, several
quantities, including two types of correlators, are investigated, leading to numerical evidence of a Berezinskii-Kosterlitz-Thouless phase transition. Exploratory results for real-time dynamics pertaining to this transition, obtained
using the approaches of variational uniform matrix product state and time-dependent
variational principle, are also discussed.}

\FullConference{37th International Symposium on Lattice Field Theory - Lattice2019\\
                16-22 June 2019\\
                Wuhan, China}

\begin{document}
\section{Introduction}
\label{sec:introduction}
Tensor-network (TN) methods, combined with lattice regularisation,
have been applied to study quantum field theories (QFTs) in recent
years~\cite{Banuls:2019rao}.    These methods allow one to work
directly with the Hamiltonian operator, without relying on Monte
Carlo simulations for path integrals.  Therefore, they offer opportunities for
solving the sign problem, and for examining real-time dynamics.

This article presents the status of our research programme of implementing
the TN strategy for the investigation of the massive Thirring model in
1+1 dimensions.  In particular, we resort to the formulation of matrix
product states (MPS) in this work.  The action of the field theory is
\begin{equation} 
\label{eq:Thirring_model_action}
    S_{\mathrm{Th}}[\psi,\bar{\psi}] \
    = \int d^2x \left[ \bar{\psi}i \gamma^{\mu}\partial_{\mu}\psi \
      - m\,\bar{\psi}\psi \
      -\frac{g}{2} \left( \bar{\psi}\gamma_{\mu}\psi \right)\left( \bar{\psi}\gamma^{\mu}\psi \right) \right] \,,
\end{equation}
with $m$ and $g$ being the fermion mass and the 
four-fermion coupling constant.  In
Refs.~\cite{Coleman:1974bu,Mandelstam:1975hb}, it was demonstrated
that the sector of zero total fermion number in this model is dual to
the sine-Gordon (SG) theory, which is also known to be connected to
the two-dimensional classical XY model~\cite{Jose:1976wc}.   This
means that a Berezinskii-Kosterlitz-Thouless (BKT)
phase transition can exist in the Thirring model.
According to perturbative analysis of the renormalisation-group (RG) flows~\cite{Amit:1979ab}, this transition
occurs at a critical coupling $g_{\ast}$ with its value being
$m-$dependent.   It can be shown in perturbation theory that $g_{\ast}
= \bar{g}_{\ast}= -\pi/2$ at $m=0$, and
$g_{\ast}$ decreases with increasing $m$.   In the regime $g < g_{\ast}$, the theory in Eq.~(\ref{eq:Thirring_model_action}) is
expected to be at criticality.  Combining with the well-known fact that the (1+1)-dimensional
massless Thirring model is a conformal field theory, one concludes
that on the $g-m$ plane, there is a fixed line, $m=0$.  This fixed
line is divided into two parts, with $(g > -\pi/2, m=0)$ being
unstable and $(g < -\pi/2, m=0)$ being stable.

In this contribution to the proceedings of the Lattice 2019 conference, we
report concluding results from our study, using the MPS strategy, for
equilibrium zero-temperature phase structure
of the massive Thirring model in 1+1 dimensions with staggered
regularisation.   These results have been published in
Ref.~\cite{Banuls:2019hzc}.   Some aspects of this investigation were
also presented at the Lattice 2017 and 2018 conferences~\cite{Banuls:2017evv, Banuls:2018ckt}.   In
addition, we describe our on-going efforts in examining real-time
dynamics 
in the massive Thirring
model in Sec.~\ref{sec:real_time}.

\section{Formulation and strategy}
\label{sec:formulation}
The classical action in Eq.~(\ref{eq:Thirring_model_action}) is
suitable for performing calculations employing the path-integral
formalism.  However, in the MPS approach, one works with the
Hamiltonian in the operator formulation of the theory.  
For this purpose, effects from the anomalous breaking of the
vector symmetry in two dimensions have to be taken into
account~\cite{Schwinger:1962tp}.   The Hamiltonian operator can 
be constructed through studying the commutation relations that are
satisfied by the energy-momentum tensor~\cite{hagen1967new}, and then be discretised on a one-dimensional lattice
using the staggered regularisation~\cite{Banks:1975gq,
  Susskind:1976jm}.   For convenience of carrying out numerical
computations, in this work the fermionic degrees of freedom in this
Hamiltonian are mapped onto spin operators through the Jordan-Wigner
(JW) transformation.   Details of the above procedure can be found in
Ref.~\cite{Banuls:2019hzc}.   Implementation of this strategy turns
the Hamiltonian operator of the continuum Thirring model, $H_{{\mathrm{Th}}}$, into that of the XXZ
spin chain coupled to both uniform and staggered magnetic fields, 
\bea
\label{eq:H_XXZ__to_sim}
 &&H_{{\mathrm{XXZ}}} = \frac{\nu (g)}{a} \bar{H}_{{\mathrm{sim}}} \,
 , \mbox{ }\mbox{ }
{\mathrm{with}}\mbox{ }\mbox{ }
 \bar{H}_{{\mathrm{sim}}} = -\frac{1}{2} \sum_{n=0}^{N-2} \left( S_{n}^{+}S_{n+1}^{-} 
            + S_{n+1}^{+}S_{n}^{-} \right)+ a \tilde{m}_{0} \sum_{n=0}^{N-1} \left(-1\right)^{n} \left(
              S_{n}^{z}+\frac{1}{2} \right)  \nonumber\\
           && \mbox{ }\mbox{ }\mbox{ }\mbox{ }\mbox{ }\mbox{ }\mbox{
           }\mbox{ }\mbox{ }\mbox{ }\mbox{ }\mbox{ }\mbox{ }\mbox{
           }\mbox{ }\mbox{ }\mbox{ }\mbox{ }\mbox{ }\mbox{ }\mbox{
           }\mbox{ }\mbox{ }\mbox{ }\mbox{ }\mbox{ }\mbox{ }\mbox{ }\mbox{ }\mbox{ }\mbox{ }\mbox{ }\mbox{ }\mbox{ }\mbox{ }\mbox{ }\mbox{ }\mbox{ }\mbox{ }\mbox{ }\mbox{ }\mbox{ }\mbox{ }\mbox{ }\mbox{ }\mbox{ }\mbox{ }\mbox{ } \mbox{ }\mbox{ }\mbox{ }\mbox{ }\mbox{ }\mbox{ }\mbox{ }\mbox{ }\mbox{ }\mbox{ }\mbox{ }\mbox{ }\mbox{ }\mbox{ }\mbox{ }\mbox{ }\mbox{ }\mbox{ }\mbox{ }\mbox{ }\mbox{ }\mbox{ }\mbox{ }\mbox{ }\mbox{ }\mbox{ }\mbox{ }\mbox{ }\mbox{ }\mbox{ }\mbox{ }\mbox{ }
        + \Delta (g) \sum_{n=0}^{N-1} \left( S_{n}^{z}+\frac{1}{2} \right) \ 
            \left( S_{n+1}^{z}+\frac{1}{2} \right) \,,
\eea
where $a$ is the lattice spacing, $N$ is the total number of lattice
sites, $S_{n}^{\pm} = S_{n}^{x} \pm i
S_{n}^{y}$ and $S_{n}^{z}$ are the spin matrices ($S^{i}_{n} =
\sigma^{i}/2$ with $\sigma^{i}$ being the Pauli matrices)
at the $n{-}$th site, and $[S^{i}_{n}, S^{j}_{m}]_{n\not=m} = 0$.  The functions $\nu (g)$ and $\Delta (g)$ are the lattice
version of wavefunction renormalisation and the four-fermion coupling~\cite{Luther:1976mt}, 
\beq
\label{eq:nu_and_Delta}
 \nu (g) = \left ( \frac{\pi - g}{\pi} \right )/ \sin\left (
 \frac{\pi - g}{2} \right ) \, , \mbox{ }\mbox{ }
   \Delta (g) = \cos \left (
 \frac{\pi - g}{2} \right )  \, , 
\eeq
and $\tilde{m}_{0} = m_{0}/\nu (g)$ with $m_{0}$ being the bare
counterpart of the mass parameter, $m$, in Eq.~(\ref{eq:Thirring_model_action}).
Since the $z-$component of the total spin corresponds to the total
fermion number in the Thirring model, the Hamiltonian that we actually use in the simulations is
\beq
\label{eq-penalty-term}
    \bar{H}_{{\mathrm{sim}}}^{{\mathrm{penalty}}} = \bar{H}_{{\mathrm{sim}}} + \lambda \left( \sum_{n=0}^{N-1} S_{n}^{z} \right)^2 \,,
\eeq
which, upon choosing $\lambda$ to be large enough (100 in this work),  ensures that the ground state obtained {\it via} a variational
search is in the sector of vanishing total
$S^{z}$~\cite{Banuls:2013jaa}.   This enables us to interpret our
results in terms of the dual SG theory and the XY model.

\section{Numerical results for the phase structure of the massive
  Thirring model}
\label{sec:phase_structure}
To study the zero-temperature phase structure of the Thirring model,
we scan the phase space by performing simulations with the
Hamiltonian, $\bar{H}_{{\mathrm{sim}}}^{{\mathrm{penalty}}}$, in
Eq.~(\ref{eq-penalty-term}) at twenty-four values of the four-fermion
coupling straddling the range $-0.9 \leq \Delta (g) \leq 1.0$, and at
$a\tilde{m}_{0} = $ 0, 0.005, 0.01, 0.02, 0.03, 0.04,
0.06, 0.08, 0.1, 0.13, 0.16, 0.2, 0.3, 0.4.   Four system sizes, $N = 400,
600, 800, 1000$, are used.  The search for the ground state is
carried out with seven choices of the bond dimension, $D = 50, 100, 200,
300, 400, 500, 600$.  The target precision of this search for the
ground-state energy is $10^{-7}$ in lattice units.   The matrix
product operator (MPO) for
$\bar{H}_{{\mathrm{sim}}}^{{\mathrm{penalty}}}$, as well as details of
our numerical implementation and analysis, can be found in Ref.~\cite{Banuls:2019hzc}.

As already reported at the Lattice
2018 conference~\cite{Banuls:2018ckt}, the von Neumann entanglement
entropy, extracted by dividing the system of size $N$ into two
subsystems between sites $n$ and $n+1$, is found to exhibit the
conformal scaling behaviour~\cite{Calabrese:2004eu} at
$a\tilde{m}_{0}=0$.  This scaling behaviour is also seen for
$a\tilde{m}_{0}\not=0$ when $g$ is smaller than a
$a\tilde{m}_{0}-$dependent value, $g_{\ast}$.

To further probe the phase structure, in the past year we performed a detailed
investigation of two types of correlation functions, namely the connected
density-density, $\la\bar{\psi}(x)\psi(x)\bar{\psi}(0)\psi(0)\ra_{{\mathrm{c}}}$, and the fermion-antifermion,
$\la\bar{\psi}(x) \psi(0)\ra$, correlators.  These correlation
functions can be written in terms of spin variables using the
staggered-fermion discretisation and the JW
transformation,
\bea
\label{eq:correlators_JW}
 \la\bar{\psi}(x)\psi(x)\bar{\psi}(0)\psi(0)\ra_{{\mathrm{c}}}
 &\longrightarrow& C_{zz}(x) =
 \frac{1}{N_{x}} \sum_{n} \left [ \la S^{z}_{n} S^{z}_{n+x} \ra - \la
   S^{z}_{n} \ra \la S^{z}_{n+x} \ra \right ] \,, \nonumber\\
 \la\bar{\psi}(x) \psi(0)\ra &\longrightarrow& C_{{\mathrm{string}}}(x)  =
 \frac{1}{N_{x}} \sum_{n} \la S^{+}_{n} S^{z}_{n+1} \cdots S^{z}_{n+x-1}
 S^{-}_{n+x} \ra \,,
\eea
where the sum over $n$ means that at a given value of $x$, we average
over all possible $C_{zz}(x)$ and $C_{{\mathrm{string}}}(x)$ computed
on the 200-site subchain straddling the middle of the lattice.   These
two correlators are expected to decay with $x$ as a power law when the
theory is at criticality, while the decay is (power-)exponentially cut-off in the gapped phase.   We fit $C_{zz}(x)$ and
$C_{{\mathrm{string}}}(x)$ to the ansatzes
\bea
\label{eq:C_fit_functions}
 C^{{\mathrm{pow}}}(x) &=& \beta x^{\alpha} + C \,, \nonumber\\
 C^{{\mathrm{pow-exp}}}(x) &=& B x^{\eta} A^{x} + C \,,
\eea
as well as other multi-exponential functions.   For both $C_{zz,{\mathrm{string}}}(x)$, our data show that for all values
of $\Delta(g)$ at $a\tilde{m}_{0}=0$,
as well as at $\Delta (g) \lesssim -0.7$ when $a\tilde{m}_{0}\not=0$,
the power-law and power-exponential fits are better compared to the
multi-exponential fits.   In this regime, it is observed that the
parameter $A$ in
Eq.~(\ref{eq:C_fit_functions}) is consistent with unity, and the
exponents $\alpha$ and $\eta$ are compatible with each other.

\begin{figure}[t!]
\begin{center}
\vspace{-0.5cm}
   \hspace{-0.7cm}
        \includegraphics[width=6.8cm, height=5cm]{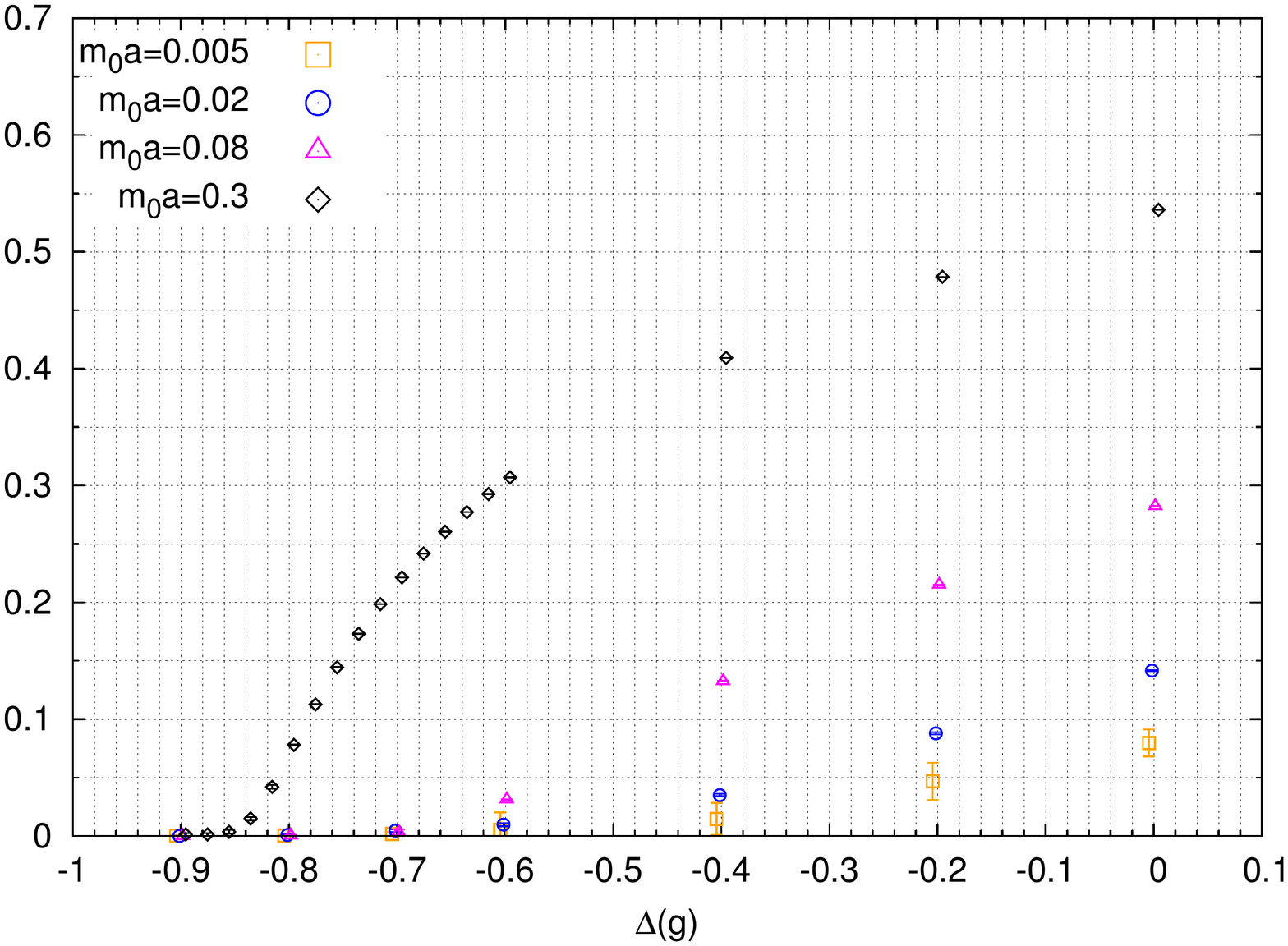}
   \hspace{0.7cm}
        \includegraphics[width=7.0cm,
        height=5cm]{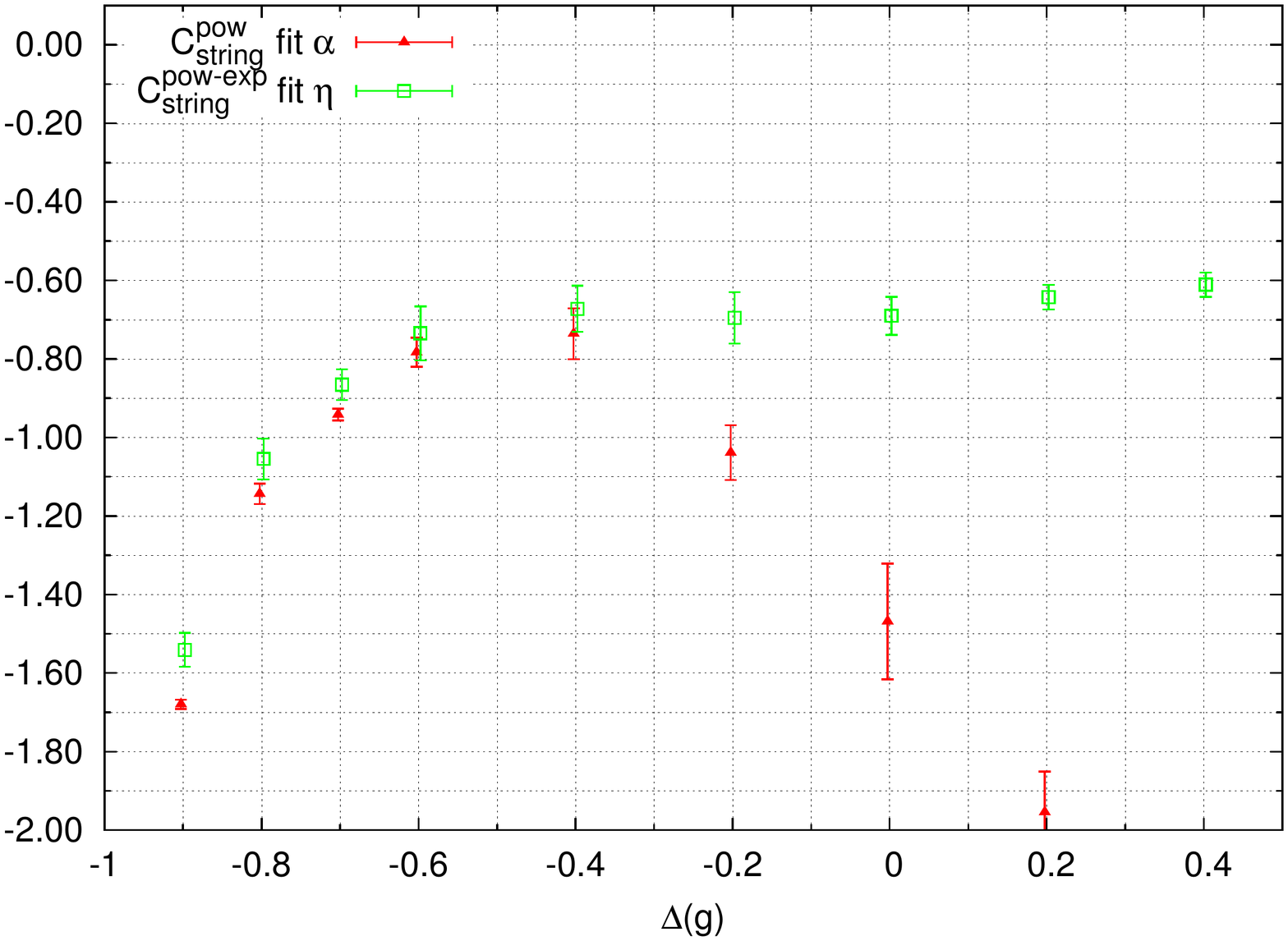}
\vspace{-0.7cm}
       \caption{Results from fitting the fermion-antifermion
         correlator, $C_{{\mathrm{string}}}(x)$, to the functions in
         Eq.~(\ref{eq:C_fit_functions}). The system size is $N=1000$. Left: Values of $C$ at
         various choices of $a\tilde{m}_{0}$  with the
         power-exponential fit in Eq.~(\ref{eq:C_fit_functions}). Right: Results of $\alpha$
         and $\eta$ introduced in Eq.~(\ref{eq:C_fit_functions}) at
         $a\tilde{m}_{0}=0.02$.  Errors in the plots are from
         systematic effects as detailed in Ref.~\cite{Banuls:2019hzc}}
\label{fig:string_corr_fits}
\end{center}
\end{figure}

It is worth noting that the constant, $C$, in Eq.~(\ref{eq:C_fit_functions}) is expected to be
consistent with zero in $C_{zz}(x)$, and our data demonstrate this
feature.   For $C_{{\mathrm{string}}}(x)$, this constant vanishes
only in the critical phase.   It is non-zero when the
fermion mass is a relevant coupling in the field theory, corresponding
to the appearance of the string order induced by the staggered
magnetic field in the spin model~\cite{Perez2008string,Bortz2007string}.  This property is also seen
in numerical results of this work.   Figure~\ref{fig:string_corr_fits} shows
examples for fits of $C_{{\mathrm{string}}}(x)$.  
From the plot on the left-hand side, it is obvious that we can
identify a region on the $\Delta(g)-a\tilde{m}_{0}$ plane where $C=0$.
This can be used to probe the phase structure.   In the right-hand
plot in Fig.~\ref{fig:string_corr_fits}, it is observed that the exponent, $\alpha$ (or
$\eta$), depends on the four-fermion coupling, $g$, in the conformal
phase, providing evidence
that the phase transition is of the BKT-type.  
It is also noted that the power-law function does not result in good fits in the gapped phase.

\begin{figure}[t!]
\begin{center}
\vspace{-0.5cm}
   \hspace{-0.7cm}
        \includegraphics[width=5.8cm,
        height=4.6cm]{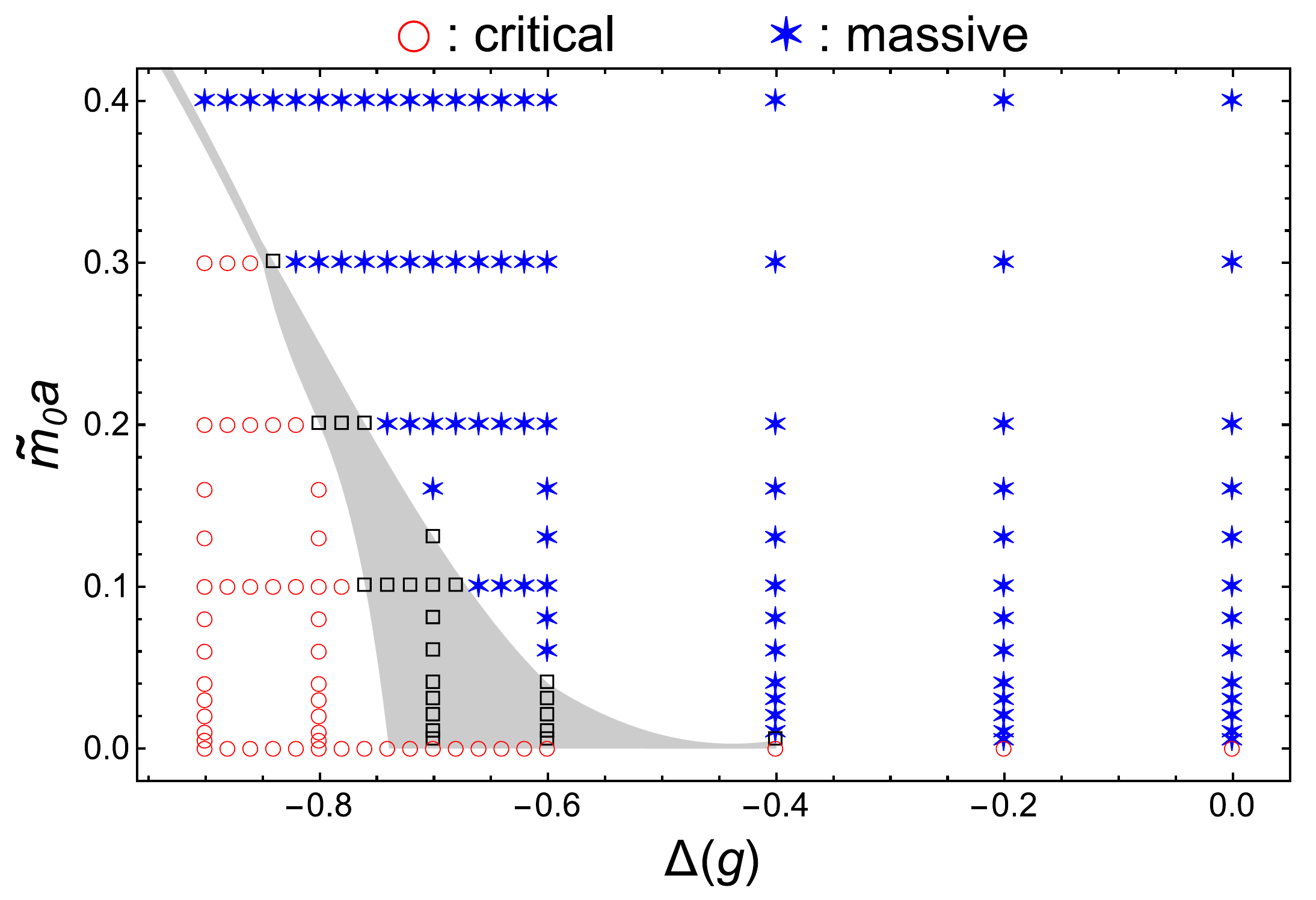}
   \hspace{0.7cm}
        \includegraphics[width=6.0cm, height=4.4cm]{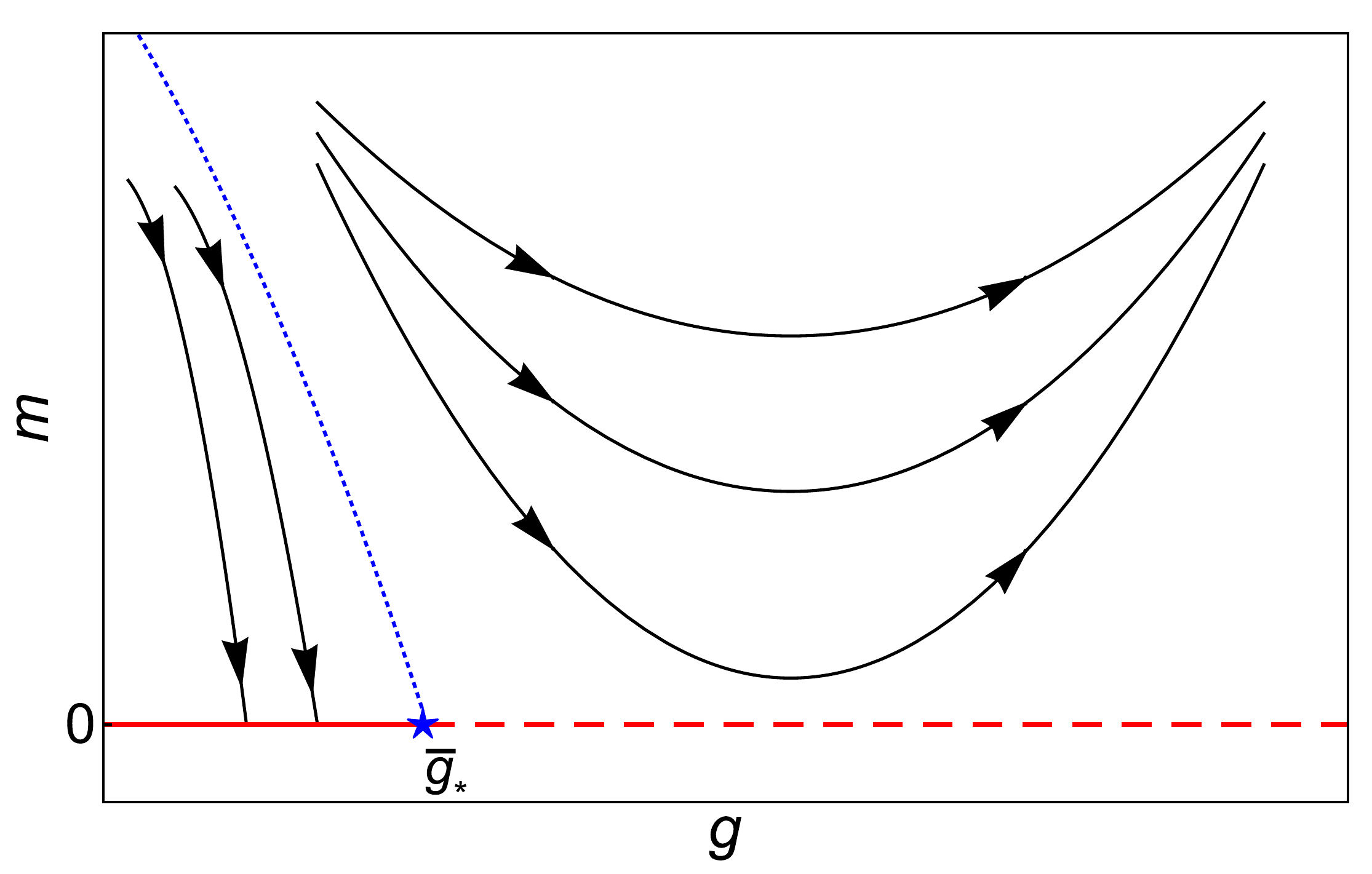}
\vspace{-0.2cm}
       \caption{Left: Results of the phase structure of the massive
         Thirring model.  Black boxes indicate points where our data
         are not precise enough to give clear answers.  Right:
         Renormalisation group flows from perturbation theory.}
\label{fig:RG_and_phase_diagram}
\end{center}
\end{figure}

Since the constant, $C$, in fitting $C_{{\mathrm{string}}}(x)$ to the
power-exponential function is the most accurately-determined 
parameter in our analysis, we use its value to probe the
phase structure of the Thirring model.  Results of this
analysis are shown in the left-hand plot in Fig.~\ref{fig:RG_and_phase_diagram}.
In the regime where $C$ is consistent with zero, it is observed that
the parameter $A$ is compatible with one.  Also displayed in the same
figure (right-hand plot) is the RG flow of the theory, obtained
using perturbative expansion in $m$ and $(g-\bar{g}_{\ast})$, with
$\bar{g}_{\ast} = -\pi/2$.    We see that
both our numerical simulation and perturbation theory predict the
existence of a phase where the fermion mass is an irrelevant coupling.

Finally, we also compute the fermion bilinear condensate, $\chi = \la
\bar{\psi}\psi \ra$, and observe that it can be non-zero when the
theory is at criticality.  This shows that $\chi$ cannot be an order
parameter for the phase transition, giving more evidence that the
transition is of the BKT-type. 

\section{Real-time dynamics}
\label{sec:real_time}
Our work on the phase structure of the model enables the investigation
of real-time dynamics pertaining to ``quenching'' across the phase
boundary.   Regarding this aspect of the study, we exploit
translational invariance in the thermodynamic limit. 
For one-dimensional systems, this allows for expressing an
infinite-size quantum state as a uniform MPS (uMPS), which can be
represented with
one bulk tensor, $A^{i}_{l,r}$,  that contains one physical index ($i$) and two
bond-dimension indices ($l$ and $r$), together with appropriate boundary tensors
when computing amplitudes and matrix elements~\cite{phien2012infinite,Zauner-Stauber:2018kxg}.
The ground state is
then extracted using the variational algorithm introduced in
Ref.~\cite{Zauner-Stauber:2018kxg}.  For the real-time evolution of
this infinite one-dimensional system, we
resort to the method of time-dependent variational principle (TDVP)~\cite{Haegeman:2011zz}.

In probing the dynamical quantum phase transition (DQPT), we compute the return rate~\cite{Heyl:DQPT2013PRL},
\beq
\label{eq:return_rate}
 G_{{\mathrm{return}}} (t) = -\lim_{N\rightarrow\infty} \frac{1}{N} {\mathrm{ln}}
 \left ( | \la 0_{1} | {\mathrm{e}}^{-i H(a\tilde{m}^{(2)}_{0}, g^{(2)}) t} | 0_{1}
   \ra |^{2} \right ) \, ,
\eeq
where $H(a\tilde{m}^{(2)}_{0}, g^{(2)})$ denotes the Hamiltonian,
$\bar{H}^{{\mathrm{penalty}}}_{{\mathrm{sim}}}$, in Eq.~(\ref{eq-penalty-term}) with the values of
the couplings set to $a\tilde{m}_{0} = a\tilde{m}^{(2)}_{0}$ and $g =
g^{(2)}$.  The state, $|0_{1}\ra$, is the vacuum of the Hamiltonian
$H(a\tilde{m}^{(1)}_{0}, g^{(1)})$.   The return rate can be extracted
by examining the spectrum of the 
``bulk transfer matrix'', 
\beq
\label{eq:transfer_matrix_def}
 \includegraphics[width=5.5cm,
 height=1.7cm]{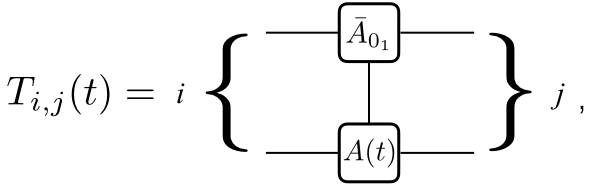}
\eeq
obtained by contracting the physical indices of the bulk tensors
$\bar{A}_{0_{1}}$ and $A(t)$.  The tensor $\bar{A}_{0_{1}}$ is the uMPS
representation of $\la 0_{1}|$, while $A(t)$ is that of exp$[-iH(a\tilde{m}^{(2)}_{0}, g^{(2)})t]|0_{1}\ra$.
Since the amplitude, $\la 0_{1}|$ exp$[-iH(a\tilde{m}^{(2)}_{0}, g^{(2)})t]|0_{1}\ra$,  is obtained through infinite
repetition of this transfer matrix, it is obvious that $G_{{\mathrm{return}}}
(t)$ can be determined using the largest eigenvalue
of $T_{i,j}(t)$.

\begin{figure}[t!]
\begin{center}
\vspace{-0.5cm}
   \hspace{-0.7cm}
        \includegraphics[width=6.5cm,
        height=4.5cm]{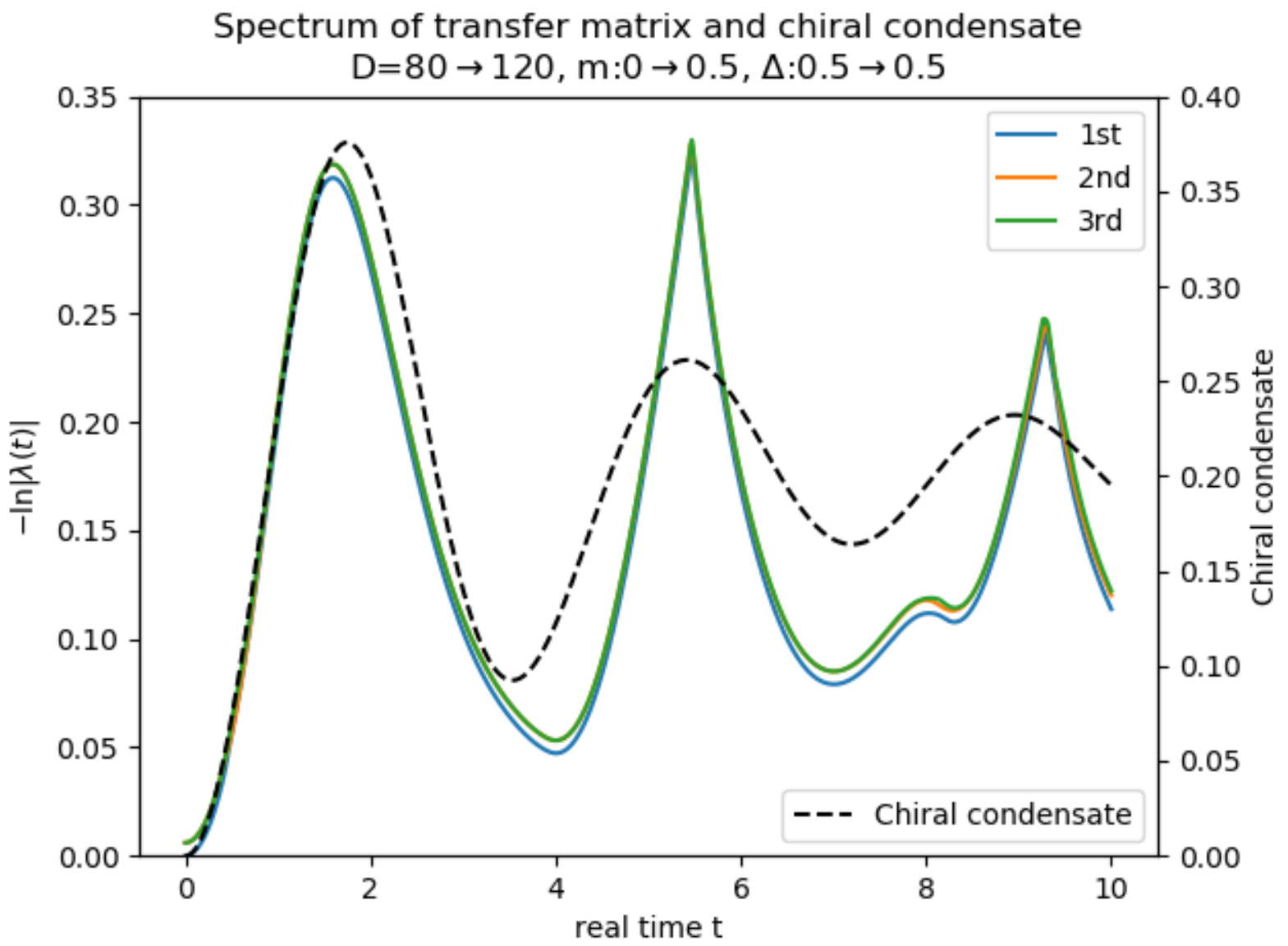}
   \hspace{0.5cm}
        \includegraphics[width=6.5cm, height=4.5cm]{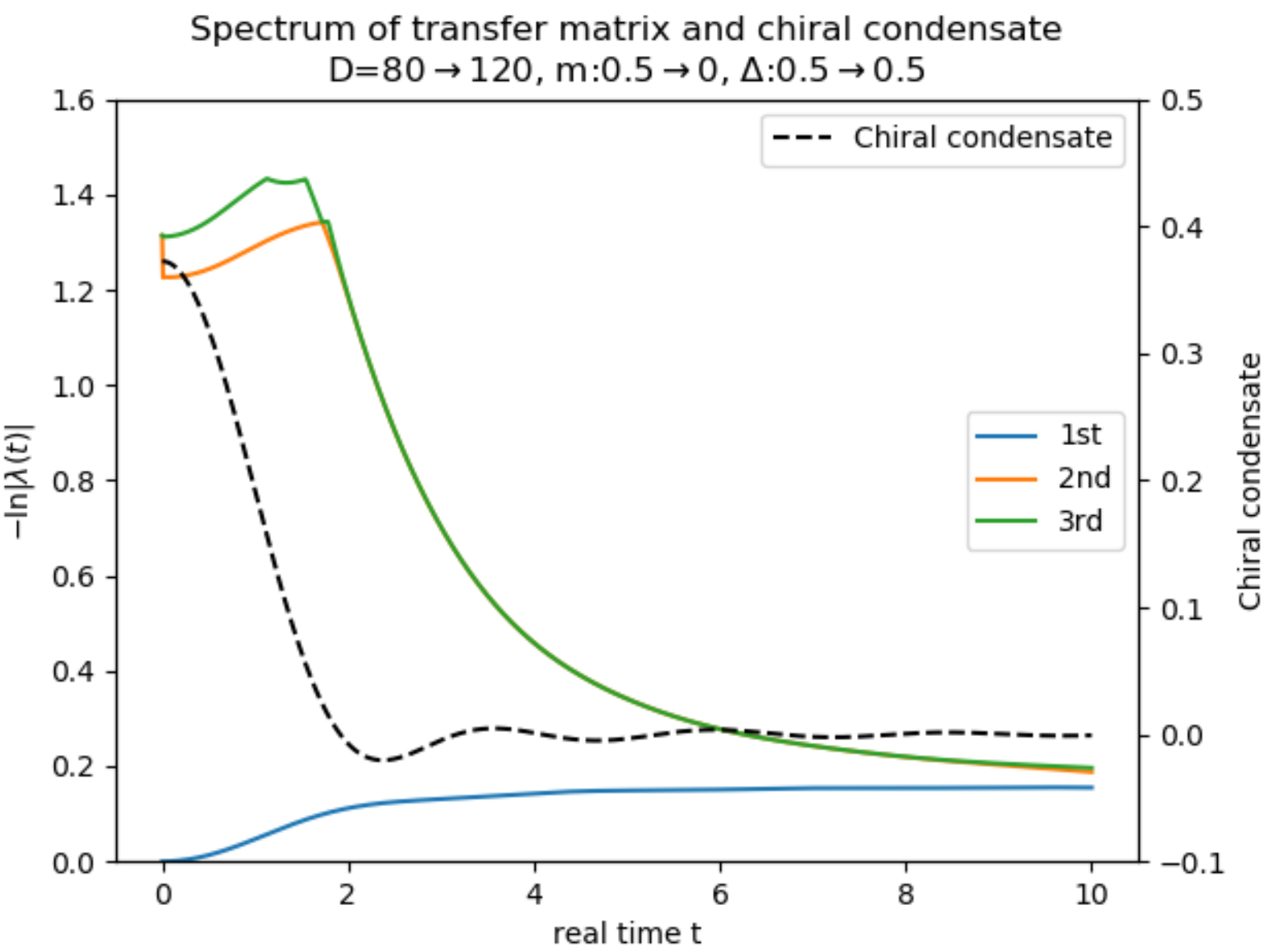}
\vspace{-0.2cm}
       \caption{Real-time evolution of the largest three
         eigenvalues (generically denoted by $\lambda(t)$) of $T_{i,j}$.  Left: Quenching from criticality to the
       gapped phase.  Right: Quenching from the gapped phase
         to
       the conformal limit.}
\label{fig:T_eigen_two_ways}
\end{center}
\end{figure}

\textcolor{black}{Dynamical quantum phase transitions are identified by non-analytic
behaviour of $G_{{\mathrm{return}}}(t)$ (or the largest eigenvalue of
$T_{i,j}(t)$ as explained above)}~\cite{Heyl:DQPT2013PRL}.
Figure~\ref{fig:T_eigen_two_ways} displays exploratory results of our
study for such transitions in the massive Thirring model, using uMPS representation of ground states at bond
dimension $D=80$.  \textcolor{black}{Numerical implementation is carried out with the
Uni10 library~\cite{Uni10}.}
We monitor the entanglement entropy, $S$, along the
real-time evolution, and increase $D$ (with an upper bound $D=120$) when $S$ exhibits signs of being saturated.
In these plots, the ``chiral condensate'' is
computed by sandwiching the JW-transformed $\bar{\psi}\psi$ operator
with real-time evolved states that are initially at $|0_{1}\ra$.
From the left-hand plot in Fig.~\ref{fig:T_eigen_two_ways}, DQPTs are observed when $|0_{1}\ra$
is the ground state of $H[a\tilde{m}^{(1)}_{0}, \Delta(g^{(1)})] =
H(0, 0.5)$, and the real-time evolution is performed using 
$H[a\tilde{m}^{(2)}_{0}, \Delta(g^{(2)})] = H(0.5,0.5)$.  That is,
DQPTs can occur while quenching from criticality into the gapped
phase.   Nevertheless, when
quenching from the massive into the conformal limit, no DQPT is seen in this
work hitherto (right-hand plot of Fig.~\ref{fig:T_eigen_two_ways}).  A
similar scenario was also noted in condensed matter physics~\cite{Andraschko:2014DQPT}.

\begin{figure}[t!]
\begin{center}
\vspace{-0.5cm}
   \hspace{-0.7cm}
        \includegraphics[width=5.5cm, height=4.5cm]{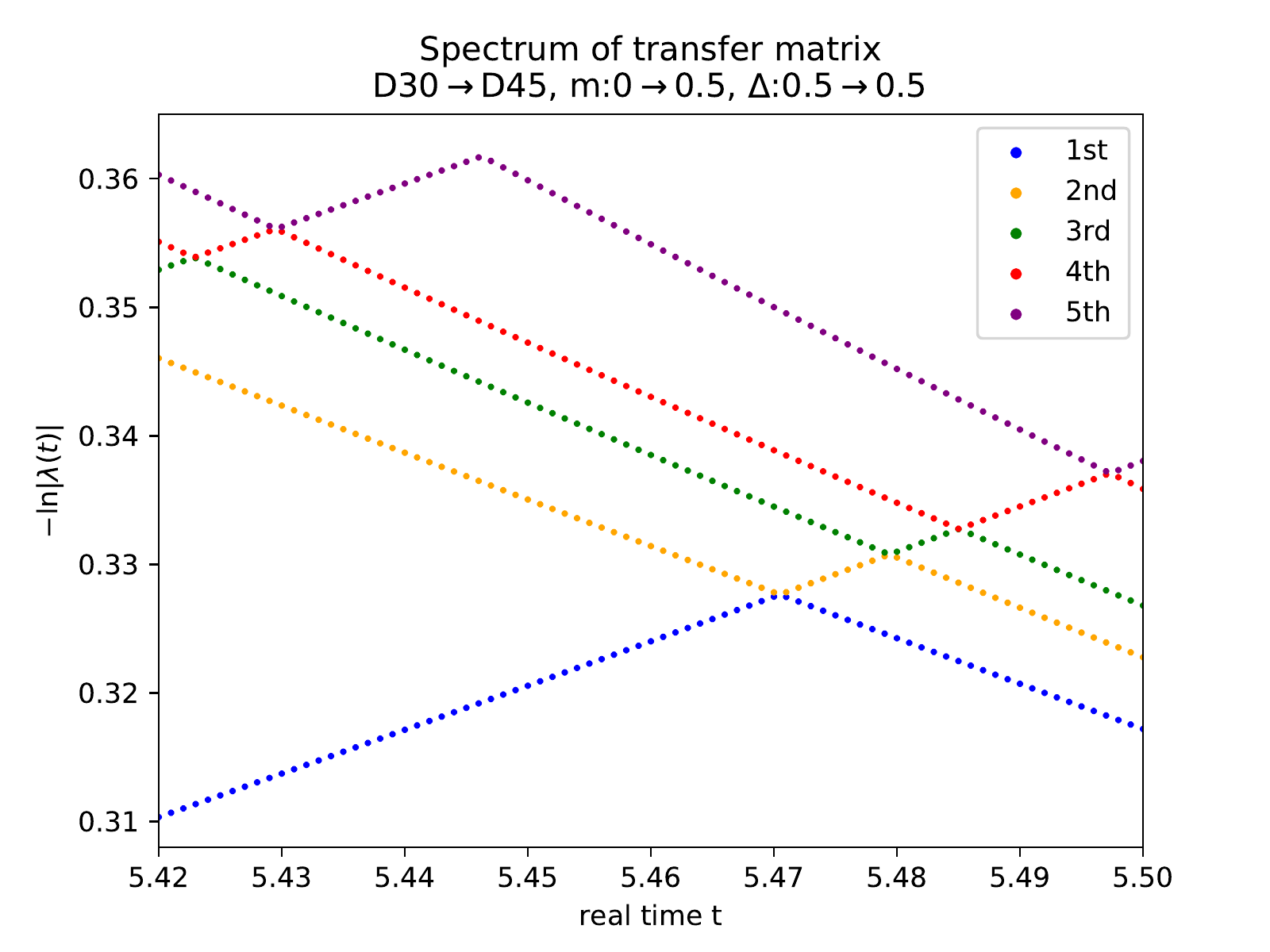}
   \hspace{-0.6cm}
        \includegraphics[width=5.5cm, height=4.5cm]{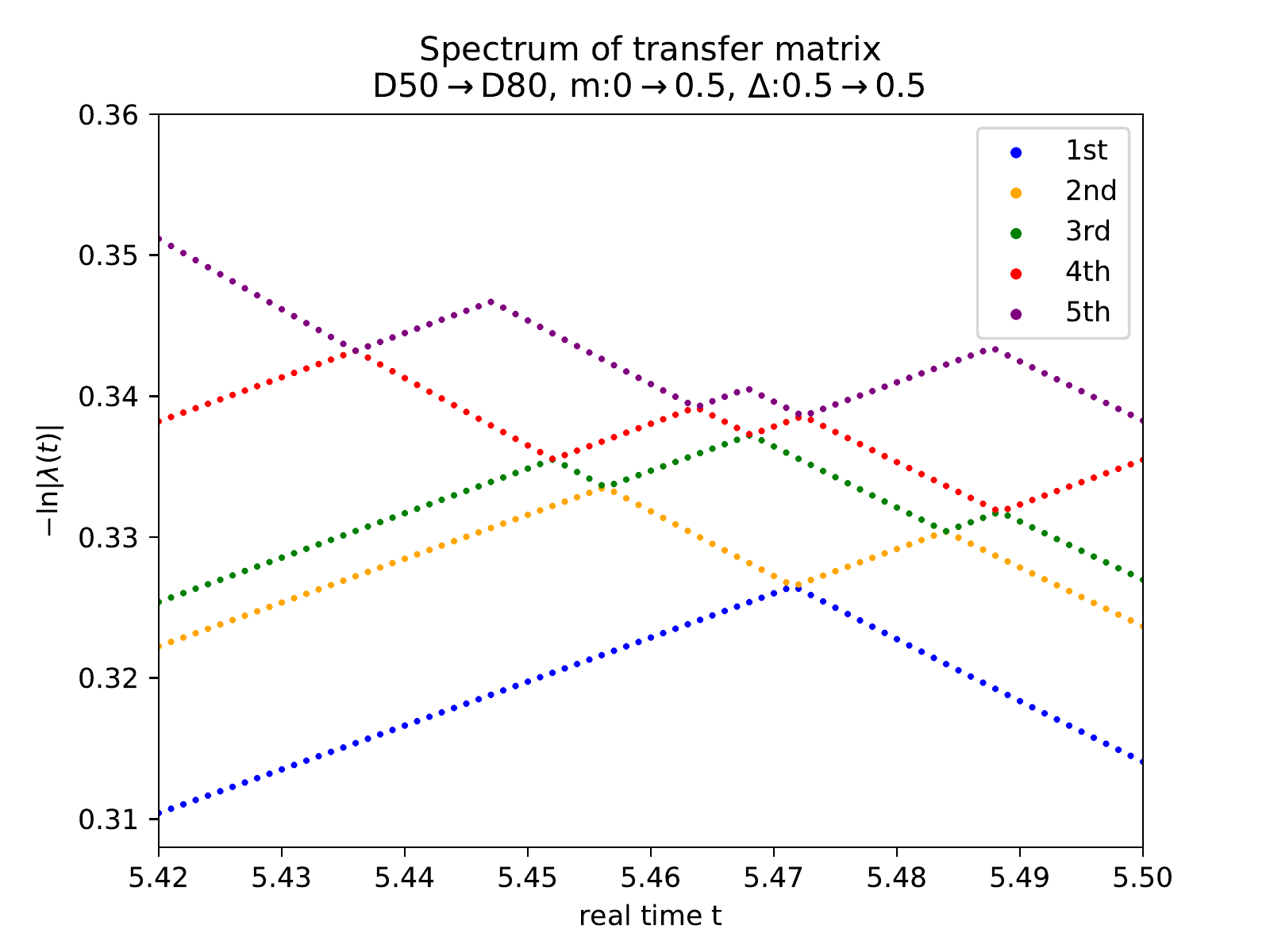}
   \hspace{-0.6cm}
        \includegraphics[width=5.5cm, height=4.5cm]{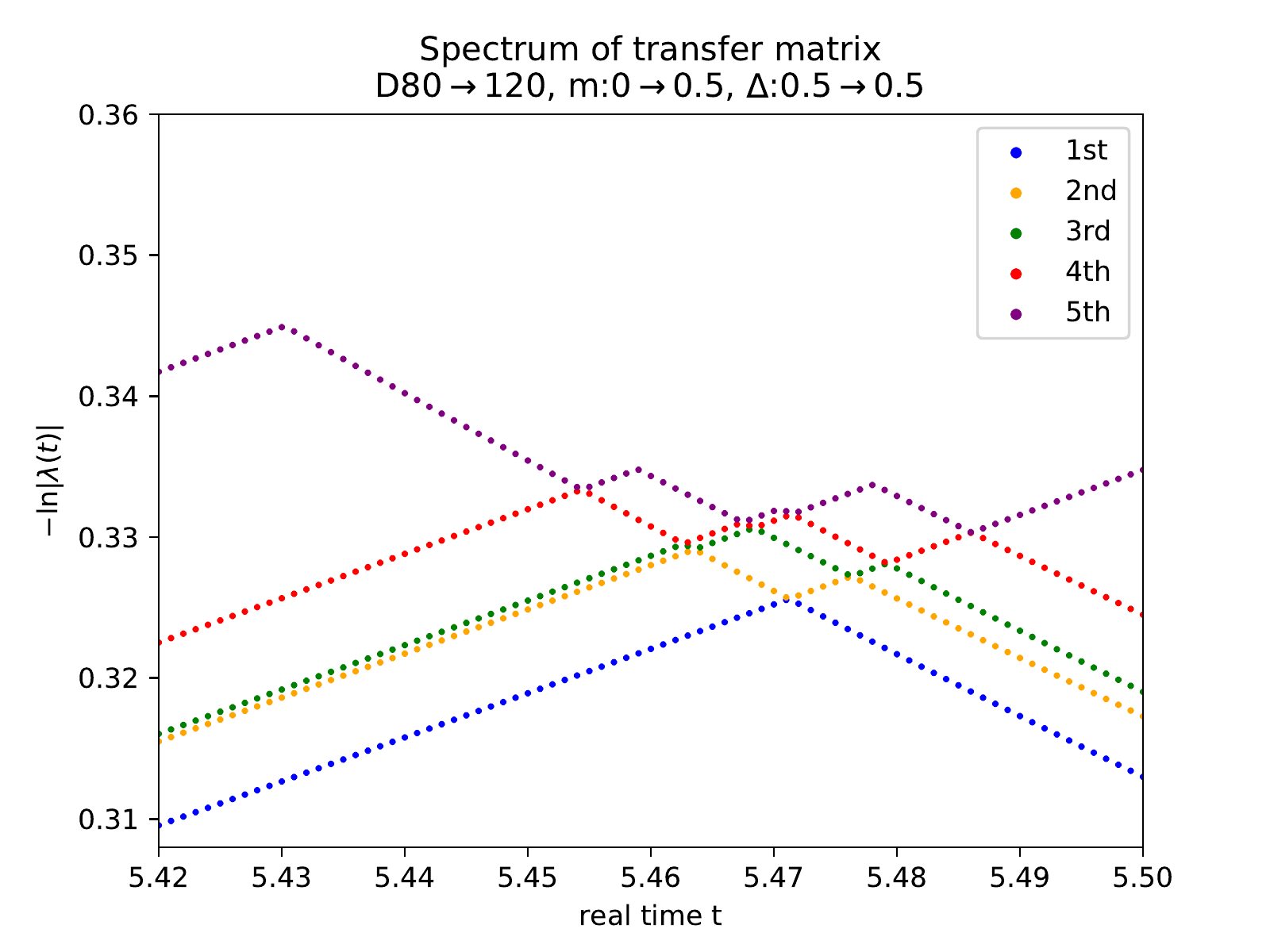}
\vspace{-0.2cm}
       \caption{Details of real-time evolution of the largest 5
         eigenvalues (generically denoted by $\lambda(t)$) of $T_{i,j}$, around a DQPT.   The 3 plots correspond
       to different choices of bond dimension.}
\label{fig:T_eigen_details}
\end{center}
\end{figure}

We examine further details of the observed DQPTs by extracting the 
largest five eigenvalues of $T_{i,j}$ around a transition point.  Results
of this study are displayed in
Fig.~\ref{fig:T_eigen_details}.  
It is observed that this DQPT occurs when level-crossing between the
largest and the second-largest eigenvalues happens.  A viable physical
picture of this phenomenon is currently being investigated.

\section{Conclusion and outlook}
\label{sec:conclusion}
\textcolor{black}{This article reports final results of our MPS study for
the zero-temperature phase structure of the massive Thirring model in
1+1 dimensions.  It is demonstrated that the approach is applicable for
probing the BKT phase transition in the theory.   We also show
exploratory results for real-time dynamics in the model.
Further numerical exploration for DQPTs, including those associated with the BKT phase
transition that occurs at fixed, non-vanishing $a\tilde{m}_{0}$, is now being carried out.}



\begin{thebibliography}{99}
%
\bibitem{Banuls:2019rao}
  M.~C.~Ba\~{n}uls and K.~Cichy,
  arXiv:1910.00257 [hep-lat].
%
\bibitem{Coleman:1974bu}
  S.~R.~Coleman,
  Phys.\ Rev.\ D {\bf 11} (1975) 2088.
%
\bibitem{Mandelstam:1975hb}
  S.~Mandelstam,
  Phys.\ Rev.\ D {\bf 11} (1975) 3026.
%
\bibitem{Jose:1976wc}
  J.~Jose,
  Phys.\ Rev.\ D {\bf 14} (1976) 2826.
%
\bibitem{Amit:1979ab}
  D.~J.~Amit, Y.~Y.~Goldschmidt and G.~Grinstein,
  J.\ Phys.\ A {\bf 13} (1980) 585.
%
\bibitem{Banuls:2019hzc}
  M.~C.~Ba\~{n}uls {\it et al.},
  Phys.\ Rev.\ D {\bf 100} (2019) 094504
  [arXiv:1908.04536 [hep-lat]].
%
\bibitem{Banuls:2017evv}
  M.~C.~Ba\~{n}uls {\it et al.},
  EPJ Web Conf.\  {\bf 175} (2018) 11017
  [arXiv:1710.09993 [hep-lat]].
%
\bibitem{Banuls:2018ckt}
  M.~C.~Ba\~{n}uls {\it et al.},
  PoS LATTICE {\bf 2018} (2018) 229
  [arXiv:1810.12038 [hep-lat]].
%
\bibitem{Schwinger:1962tp}
  J.~S.~Schwinger,
  Phys.\ Rev.\  {\bf 128} (1962) 2425.
%
\bibitem{hagen1967new}
  C.~R.~Hagen,
  Nuovo Cimento {\bf B51} (1967) 169.
%
\bibitem{Banks:1975gq}
  T.~Banks, L.~Susskind and J.~B.~Kogut,
  Phys.\ Rev.\ D {\bf 13} (1976) 1043.
%
\bibitem{Susskind:1976jm}
  L.~Susskind,
  Phys.\ Rev.\ D {\bf 16} (1977) 3031.
%
\bibitem{Luther:1976mt}
  A.~Luther,
  Phys.\ Rev.\ B {\bf 14} (1976) 2153.
%
\bibitem{Banuls:2013jaa}
  M.~C.~Ba\~{n}uls {\it et al.},
  JHEP {\bf 1311} (2013) 158
  [arXiv:1305.3765 [hep-lat]].
%
\bibitem{Calabrese:2004eu}
  P.~Calabrese and J.~L.~Cardy,
  J.\ Stat.\ Mech.\  {\bf 0406} (2004) P06002
  [hep-th/0405152].
%
\bibitem{Perez2008string}
 D.~Perez-Garcia {\it et al.},
 Phys.\ Rev. Lett. {\bf 100(16)} (2008) 167202
 [arXiv:0802.0447 [cond-mat.str-el]].
%
\bibitem{Bortz2007string}
 M.~Bortz {\it et al.},
 J.\ Phys.\ A {\bf 40} (2007) 4253
 [arXiv:cond-mat/0612348].
%
%
\bibitem{phien2012infinite}
 H.~N.~Phien {\it et al.},
 Phys.\ Rev.\ B {\bf 86} (2012) 245107
 [arXiv:1207.0652 [quant-ph]].
%
%
\bibitem{Zauner-Stauber:2018kxg}
  V.~Zauner-Stauber {\it et al.},
  Phys.\ Rev.\ B {\bf 97} (2018) no.4,  045145
  [arXiv:1701.07035 [quant-ph]].
%
\bibitem{Haegeman:2011zz}
  J.~Haegeman {\it et al.},
  Phys.\ Rev.\ Lett.\  {\bf 107} (2011) 070601
  [arXiv:1103.0936 [cond-mat.str-el]].
%
\bibitem{Heyl:DQPT2013PRL}
  M.~Heyl {\it et al.},
  Phys.\ Rev.\ Lett. {\bf 110} (2013) 135704
  [arXiv:1206.2505[cond-mat.stat-mech]].
%
\bibitem{Uni10}
 Y.-J.~Kao, Y.-D.~Hsieh and P.~Chen,
 J.\ of\ Phys.\ Conf.\ Series\ {\bf 640} (2015) 012040.
%
\bibitem{Andraschko:2014DQPT}
  F.~Andraschko and J.~ Sirker,
 Phys.\ Rev.\  B {\bf 89} (2014)125120
 [arXiv:1312.4165 [cond-mat.str-el]].
%
\end{thebibliography}
\end{document}